# Emergence of Tertiary Dirac Points in Graphene Moiré Superlattices

Guorui Chen[1,2], Mengqiao Sui[1,2], Duoming Wang[3], Shuopei Wang[3], Jeil Jung[4], Pilkyung Moon[5], Shaffique Adam[6], Kenji Watanabe[7], Takashi Taniguchi[7], Shuyun Zhou[8,9], Mikito Koshino[10], Guangyu Zhang[3] and Yuanbo Zhang[1,2*]

[1]*State Key Laboratory of Surface Physics and Department of Physics, Fudan University, Shanghai 200433, China*

[2]*Collaborative Innovation Center of Advanced Microstructures, Fudan University, Shanghai 200433, China*

[3]*Beijing National Laboratory for Condensed Matter Physics and Institute of Physics, Chinese Academy of Sciences, Beijing 100190, China.*

[4]*Department of Physics, University of Seoul, Seoul 02504, Korea.*

[5]*New York University and NYU-ECNU Institute of Physics at NYU Shanghai, Shanghai 200120, China.*

[6] *Graphene Research Centre and Department of Physics, National University of Singapore, 2 Science Drive 3, 117551, Singapore.*

[7]*Advanced Materials Laboratory, National Institute for Materials Science, 1-1 Namiki, Tsukuba, 305-0044, Japan.*

[8]*State Key Laboratory of Low Dimensional Quantum Physics and Department of Physics, Tsinghua University, Beijing 100084, China*

[9]*Collaborative Innovation Center of Quantum Matter, Beijing, P.R. China*

[10]*Department of Physics, Osaka University, Toyonaka 560-0043, Japan*

[*]Email: zhyb@fudan.edu.cn

**The electronic structure of a crystalline solid is largely determined by its lattice structure. Recent advances in van der Waals solids, artificial crystals with controlled stacking of two-dimensional (2D) atomic films[1], have enabled the creation of materials with novel electronic structures. In particular, stacking graphene on hexagonal boron nitride (hBN) introduces moiré superlattice that fundamentally modifies graphene's band structure and gives rise to secondary Dirac points (SDPs)[2–8]. Here we find that the formation of a moiré superlattice in graphene on hBN yields new, unexpected consequences: a set of tertiary Dirac points (TDPs) emerge, which give rise to additional sets of Landau levels when the sample is subjected to an external magnetic field. Our observations hint at the formation of a hidden Kekulé superstructure on top of the moiré superlattice under appropriate carrier doping and magnetic fields.**

2D materials, first exemplified by graphene[9] and later joined by metal dichalcogenides[10–12] and phosphorene[13,14], have emerged as a new class of material that offers a wide range of material properties distinctively different from their bulk crystals. Such atomically thin crystals are entirely exposed as surface, and their electronic properties are therefore extremely sensitive to their surroundings. So when layers of different 2D materials are brought together to form van der Waals solids[1] via controlled stacking, the coupling between the layers fundamentally alters the electronic structure of the original material. Controlled stacking thus provides a route to engineering artificial materials with designer properties.

The concept of van der Waals solid has been best demonstrated in graphene supported on hBN substrate[15–18]. The lattice mismatch between graphene and hBN in a vertical heterostructure gives rise to moiré patterns that profoundly alters the electronic structure of graphene. Specifically, the moiré patterns create a periodic superlattice potential. When subjected to strong magnetic fields, the electrons in such a 2D superlattice give rise to a fractal spectrum referred to as Hofstadter's butterfly[19] – a long-standing prediction that was recently observed in graphene/hBN heterostructures[4–6]. Meanwhile, the periodic moiré patterns fold the original Dirac fermion bands of graphene into mini-Brillouin zones, and a set of secondary Dirac points (SDP) emerge at the mini-zone corners[3,5,7,8,20,21].

Here, we show that the graphene/hBN van der Waals heterostructures harbor new, unexpected electronic states. Namely, tertiary Dirac points (TDP) appear at high hole doping levels. The TDPs produce additional sets of Landau levels (LLs) under magnetic fields. The degeneracy of the LLs is found to be a factor of three larger than for the first- and second-generation Dirac points, which indicates a peculiar band topology at the TDPs. Analysis of the exact band filling at the TDPs and the Zak oscillation period suggests that the TDPs may originate from a $\sqrt{3} \times \sqrt{3} R30°$ Kekulé superstructure[22] on top of the moiré pattern. A systematic study of graphene/hBN heterostructures with varying stacking angle shows that the observed electronic structure at TDP is universal to all our samples where TDPs are accessible. Our results add to the list of novel properties of graphene/hBN heterostructures, and showcase the powerful concept of van der Waals solid engineering.

Graphene and hBN share a common honeycomb lattice structure. When graphene is placed on top of hBN, the slight mismatch between the two lattices (graphene's lattice constant is $\delta = 1.8\%$ smaller than that in hBN[23]) induces quasi-periodic patterns, i.e. moiré patterns, that form a superlattice also with a honeycomb structure. The wavelength $\lambda$ of the moiré superlattice is determined by the relative stacking angle $\phi$ between the two lattices[3,24,25]. Such a moiré superlattice induces profound modifications to the electronic structure of graphene, which is the focus of this study.

We were able to engineer electronic bands of graphene by varying $\phi$ precisely and systematically. We employ two types of samples to cover a wide range of $\phi$: 1) epitaxial graphene grown on hBN flakes using remote plasma assisted chemical vapor deposition[7] and 2) mechanically exfoliated graphene transferred on top of hBN flakes with precisely controlled stacking angles[26,27]. Optical micrographs of typical sample of type 1 and 2 are shown in the upper panels of Fig. 1a and 1b, respectively. We choose hBN flakes with thicknesses ranging from 20 to 80 nm as substrates for both types of samples. The epitaxial graphene aligns with the hBN substrate without rotation so that $\phi = 0°$ (see ref. 7 for more details of sample growth). Meanwhile $\phi$ can take any value in the type 2 graphene/hBN assembly. We identify the crystal orientation of graphene and hBN using the crystalline edges of the flakes, and manually align the graphene lattice with the hBN flake under an optical microscope. Precise control of $\phi$ can be achieved with an accuracy down to $\pm 0.5°$. Finally, metal electrodes (Cr/Au, 1 nm and 100 nm, respectively) are defined on top of the samples for subsequent electronic transport measurements (Fig. 1a and 1b). The devices are fabricated on Si wafers covered with 285 nm SiO$_2$ layer, and the degenerately doped Si is used as back gate to tune the carrier density $n = C_g V_g$ in graphene, where $V_g$ is the gate voltage and $C_g$ is the gate capacitance per unit area.

Both types of samples exhibit periodic moiré patterns that are resolved by Atomic Force Microscopy (AFM). AFM images of typical epitaxial and transferred graphene on hBN substrates are displayed in Fig. 1a and 1b, respectively. (The AFM images also give a direct measurement of the wavelength $\lambda$ of the moiré pattern[28].) The formation of a moiré superlattice alters the electronic band structure of graphene[2,8,20,29,30]. Specifically, the superlattice gives rise to mini-Brillouin zones in the reciprocal space, and the band-folding at the mini-Brillouin zone boundaries leads to mini-bands with secondary Dirac points (SDPs) at the corners of the first mini-Brillouin zone (Fig. 1c). The SDPs correspond to minima in graphene's density of states (DOS) on both electron and hole side of the gate doping, which are manifested as satellite peaks in the sample resistance $R_{xx}$ measured as a function of $n$ (marked by orange triangles, Fig. 2a). The peaks marked by black triangles at $n \approx 0$ correspond to the primary Dirac point (PDP)).

The formation of SDPs is further corroborated by zero-crossings in $R_{xy}$ under a weak magnetic field (Supplementary Figure 3). Because the sign of $R_{xy}$ is dictated by the sign of charge carriers, such zero-crossings signify the transition from electron-like carriers (positive $R_{xy}$) to hole-like carriers (negative $R_{xy}$), as the Fermi level passes the PDP (black triangles, Supplementary Figure 3) and the SDPs (orange triangles, Supplementary Figure 3). We note that the van Hove singularities associated with saddle points within the mini-bands also induces sign-reversal in $R_{xy}$ (ref. 8); the singularity was observed as zero-crossing of $R_{xy}$ between the PDP and the SDP in all samples shown in Supplementary Figure 3.

Varying the stacking angle $\phi$ leads to systematic changes in the electronic structure of graphene. As shown in Fig. 2a, an increasing $\phi$ (decreasing $\lambda$) consistently shifts the SDP to higher electron (or hole) densities. The observed $\phi$-dependence can be attributed to the formation of mini-bands in graphene in the presence of the moiré superlattice. Specifically, one such mini-band contains 4 electrons per unit cell of the superlattice, where the degeneracy of 4 comes from the two-fold spin and two-fold valley degeneracy. The SDP is, therefore, expected to appear at doping level $n_{SDP} = 4n_0$, where $n_0 = 1/A_0$ and $A_0 = \sqrt{3}\lambda^2/2$ is the unit cell area of the moiré superlattice. Fig. 2b displays the doping level at the satellite peaks in a series of samples plotted as a function of the moiré superlattice wavelength $\lambda$, which is extracted from AFM measurements of the same samples (Supplementary Figure 1). Indeed, we find that the carrier density at the satellites exhibits a $n_{SDP} \sim \lambda^{-2}$ dependence, and the model fits the data well with no adjustable parameters (Fig. 2b). Such agreement provides additional, unambiguous evidence that the peaks in $R_{xx}$ and the corresponding zero-crossings in $R_{xy}$ are manifestations of SDPs in the presence of a moiré superlattice.

Having established the link between the $R_{xx}$ satellite peak and SDP, the observation of an additional $R_{xx}$ peak at $n \approx 6.6n_0$ (marked by blue triangle on each curve in Fig. 2a) came as a surprise: a new generation of higher order Dirac points may exist beyond SDP. Indeed, the $R_{xx}$ peaks are accompanied by zero-crossings in low-field $R_{xy}$ measured as a function of carrier density. The new $R_{xx}$ peak and its corresponding $R_{xy}$ zero-crossing are observed on every sample as long as they are accessible within gate doping limit, and their location consistently shifts to higher carrier density as $\lambda$ decreases (blue triangles, Supplementary Figure 3). But here we note that the zero-crossings occur at $n \approx 8n_0$, away from $R_{xx}$ peak position. Such misalignment may be a result of complex band structure at the TDP; additional bands may shift the zero-crossing away from the corresponding peak in $R_{xx}$ (see Supplementary Figure 3 for detailed discussion).

In high magnetic fields, the formation of LLs provides insight into the electronic structure at all three generations of Dirac points. Fig. 3a displays a LL fan diagram (conductivity, $\sigma_{xx}$, recorded as a function of $n$ and magnetic field, $B$) obtained from a typical sample. Here $\sigma_{xx} = \frac{R_{xx}}{R_{xx}^2 + (w/L)^2 R_{xy}^2}$, where $w$ is channel width, and $L$ is channel length, is related to the DOS. A set of LLs forms at every generation of Dirac points; the slope of the straight line that tracks the LL yields the filling-factor $\nu$ (in unit of $eB/h$). At PDP and SDP, the filling-factors at fully filled LLs produce the sequence $\nu = g(N + \frac{1}{2})$ that is characteristic of massless Dirac fermions. Here $N$ is integer and $g = 4$ is the LL degeneracy at both PDP and SDP. $g = 4$ at the PDP reflects the two-fold spin degeneracy and the two-fold $\boldsymbol{K}/\boldsymbol{K'}$ valley degeneracy of the original graphene band structure. The same $g = 4$ degeneracy at the SDP implies that there is, surprisingly, only one SDP in the mini-Brillouin zone (mBZ) at each PDP (Fig.

3c). Such 1:1 correspondence deviates from a simple band-folding picture that predicts two SDPs per PDP[3,5], but agrees with recent angle-resolved photoemission spectroscopy (ARPES) measurements performed on the same types of samples[31] and band structure calculations[32]. The deviation highlights the important role of lattice distortion and superlattice symmetry-breaking in determining the band structure of graphene on hBN[8,20,32].

At the TDP, the appearance of a new set of LLs further corroborate the emergence of a mini-band, and here we observe a peculiar pattern in the fan diagram. The filling factors at the first three half-filled LLs are identified as $\nu = -6, -18, -30$, which follows the sequence $\nu = g(N + \frac{1}{2})$, where $g = 12$ (Fig. 3a; see Supplementary Figure 5 for detailed analysis). There are two main points to notice: i) the sequence is similar to that at PDP and SDP, indicating that the carriers at TDP are still of a Dirac fermion nature; ii) the twelve-fold degeneracy at the TDP implies that there are three TDPs in the mBZ at each PDP (Fig. 3d). No TDP with such peculiar electronic structure was reported in previous studies of graphene/hBN heterostructures.

The question then arises as to what specifically caused the TDPs. An important clue comes from analysis of the horizontal lines in the LL fan diagram that are periodic in $1/B$ – the so-called Zak oscillations[4,5] (Fig. 3a). The Zak oscillations appear when the magnetic flux penetrates the unit cell area of the superlattice in unit fractions, so the periodicity of the oscillations is described by $1/B = q/B_0$, where $q$ is integer and $B_0 = \Phi_0/A_0$ is the quantum flux, $\Phi_0$, per superlattice unit cell area, $A_0$. Measurement of the Zak period $1/B_0$, therefore, gives a measurement of the area of the superlattice unit cell from which the LLs originate. To this end, we replot the LL fan diagram in Fig. 3 as a function of $1/B$ and the filling factor $\nu$ at SDP (Fig. 4a) and TDP (Fig. 4b). The oscillations, marked by orange and black arrows in Fig. 4a and 4b, respectively, are clearly resolved. (The filling factor is calculated from $n$ using $\nu = (n - n_D)/n_L$, where $n_D = 4n_0$ at the SDP and $n_D = 6.6n_0$ at the TDP; $n_L = eB/h$ is the carrier density that can be accommodated in each LL). At the SDP, $q$ and $1/B$ extracted from each oscillation in Fig. 4a falls on a straight line (Fig. 5a), confirming the periodicity of the Zak oscillations. From the slope of the line we obtain a moiré wavelength of $\lambda = 12.9$ nm that is responsible for the SDP, in excellent agreement with an independent estimation (12.7 nm) obtained from zero-field $R_{xx}$ data taken on the same sample (second curve from top, Fig. 2a).

We now turn to the Zak oscillations at the TDP. Here the linear dependence between $q$ and $1/B$ extracted from Fig. 4b attests the periodicity of the oscillations (Fig. 5a). A wavelength of 21.5 nm extracted from the slope of the line fit, however, is about $\sqrt{3}$ times larger than the moiré pattern wavelength $\lambda$. The unit cell area of the superlattice that causes the Zak oscillations at TDP is therefore three times larger than the moiré unit-cell area $A_0$. Given that the superlattice should respect the symmetry of the underlying moiré pattern, our result suggests that the $\sqrt{3} \times \sqrt{3}R30°$ Kekulé superstructure on top of the moiré superlattice, illustrated schematically in Fig. 5b, is the origin of the LL formation at TDP. Such $\sqrt{3} \times \sqrt{3}R30°$ Kekulé superstructure will presumably induce higher order band-folding of the mini-bands of the moiré superlattice and lead to additional peaks in $R_{xx}$, which we observed under zero magnetic field.

Particularly, the superstructure with a unit-cell area of $3A_0$ suggests that the $R_{xx}$ peak at the TDP $n_{TDP}/n_0 \approx (7-1/3)$ incorporates an additional $s = -1/3$ from an otherwise integral hole filling in the Wannier diagram equation for the Hofstadter gaps $n/n_0 = s + t\,\Phi/\Phi_0$, where $\Phi$ is the magnetic flux and $t$ is an integer. In recent experiments it was noted that Hofstadter gaps[33] with fractal $s = 1/3$, $-1/3$ intercepts that arise in sufficiently strong magnetic fields ($\Phi/\Phi_0 \sim 0.8$) are compatible with electron or hole charge density wave (CDW) phases. The CDW phases are commensurate with the substrate potential, but may have a larger period[34,35]. In this scenario, the magnetic-field-dependent Coulomb interaction, $E_{Coul} \approx e^2/\varepsilon l_B$, drives the formation of the CDW phase (here $e$ is the charge of an electron and $\varepsilon \approx 4$ is the dielectric constant; the magnetic length $l_B \propto B^{-1/2}$). In our experiments a magnetic field on the order of 1 T is enough to trigger the CDW phase, so the Coulomb energy that gives rise to the Kekulé structure is on the order of 10 meV, comparable in order of magnitude to the substrate-induced potential oscillations that define the moiré superlattice template. Comparison with earlier band structure calculations for carrier densities corresponding to the TDP[32] and the fan diagrams for the non-interacting realistic Graphene/hBN Hamiltonians[36] suggest that the actual interaction-renormalized band structure is probably more complex than the isolated sets of degenerate Dirac cones located at symmetry points of the moiré Brillouin zone. A detailed analysis of such effects calls for a more comprehensive theoretical study.

**Figure captions**

**Figure 1 | Moiré superlattice in Graphene/hBN heterostructure. a-b,** Optical and AFM images of typical epitaxial (**a**) and mechanically transferred (**b**) graphene on hBN

flakes. The epitaxial graphene covers the entire hBN flake in **a**, and the transferred graphene flake is etched into Hall bar (outlined by broken line) in **b**. AFM images are false-color rendering of phase signal recorded in non-contact mode. Hexagons in lower panels mark the unit cells of the moiré superlattices. **c**, Schematic of calculated band structure of graphene moiré superlattice. Zero stacking angle is assumed. Black hexagon represents the mBZ. PDP is located at the center of the mBZ. SDP is present at K point, but absent at K' point. As each K point is shared by three neighboring mBZs, each SDP corresponds to one PDP.

**Figure 2 | Emergence of SDP and TDP in graphene moiré superlattices with varying wavelength. a**, Resistance in logarithmic scale as a function of carrier density obtained from samples with varying moiré wavelength. Curves are shifted vertically for clarity. Black, orange and blue triangles mark the resistance peaks corresponding to PDP, SDP and TDP, respectively. **b**, Carrier density at SDP plotted as a function of moiré wavelength obtained from AFM measurements. Data points are extracted from measurements in **a** and Supplementary Figure 1. The dashed line is the fit with the simple band-folding model, $n \sim \lambda^{-2}$, described in the main text. Inset: carrier density at TDP, $n_{TDP}$, plotted as a function of moiré wavelength.

**Figure 3 | LL fan diagram obtained in a graphene moiré superlattice. a**, LL fan diagram, i.e. longitudinal conductivity plotted as a function of magnetic field and normalized carrier density, $n/n_0$. White, orange and black broken lines mark the three

sets of LLs developed at PDP, SDP and TDP, respectively. Integers on each line denote the filling factors extracted from the slope of each line. **b-d**, Schematic drawing of the LL configuration at PDP (**b**), SDP (**c**) and TDP (**d**). The degeneracy at the three generation of Dirac points are in the ratio of 1:1:3.

**Figure 4 | Zak oscillations at SDP and TDP. a-b**, Longitudinal conductivity plotted as a function of $1/B$ and filling factor $v$ at SDP (**a**) and TDP (**b**). Zak oscillations at SDP and TDP are marked by arrows in **a** and **b**, respectively. Vertical broken lines indicate fully filled LLs.

**Figure 5 | $\sqrt{3} \times \sqrt{3}R30°$ Kekulé superstructure in graphene/hBN heterostructure. a**, Fan diagram of the Zak oscillations at SDP (orange) and TDP (black). Location of $1/B$ for the $q$th Zak oscillation plotted as a function of $q$. Data points are extracted from the Zak oscillations shown in Fig. 4a and 4b. Lines are linear fit to the data. From the slope of the line fits, we obtain moiré wavelength of 12.9 nm at SDP and 21.5 nm at TDP. **b**, Schematic of the moiré superlattice in graphene/hBN heterostructure. Gray circles denote carbon atoms in graphene, and red and blue circles denote boron and nitrogen atoms in hBN, respectively. The orange hexagon marks the unit cell of the original moiré superlattice, and black hexagon marks the unit cell of a presumed $\sqrt{3} \times \sqrt{3}R30°$ Kekulé superstructure on top of the moiré superlattice.

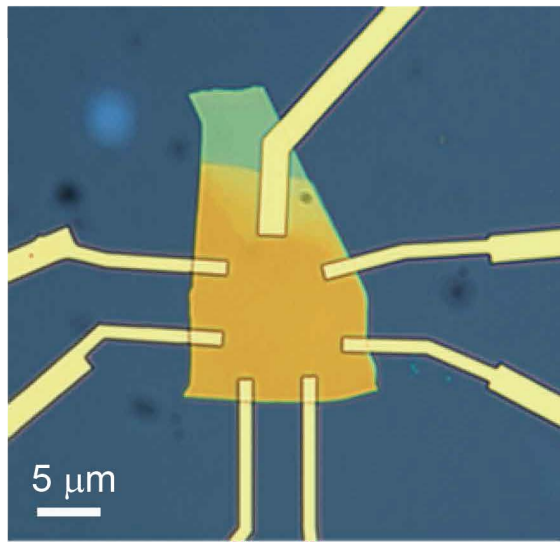 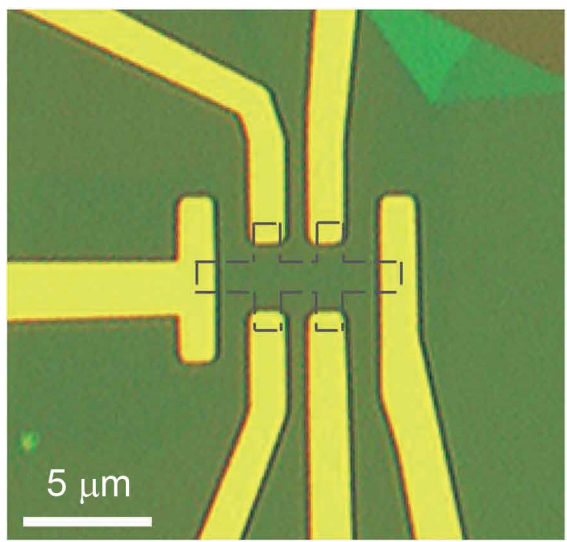 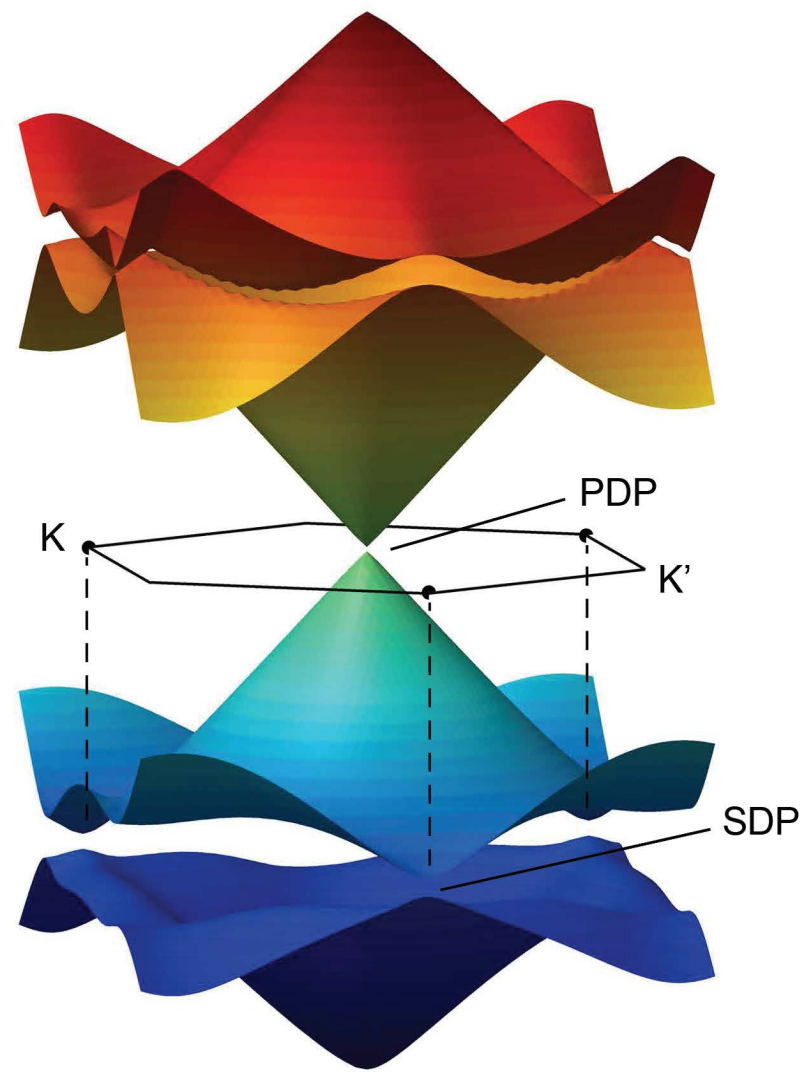
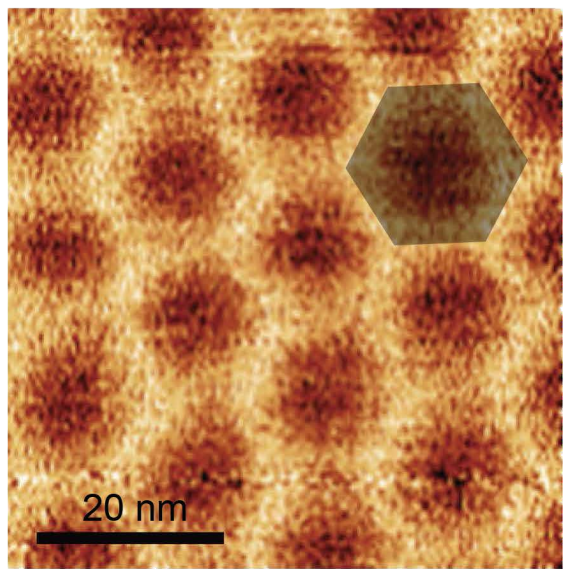 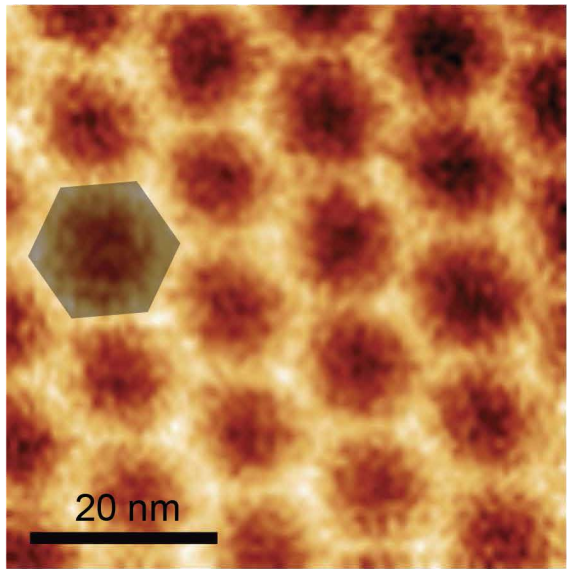

Fig. 1, G. Chen *et al*.

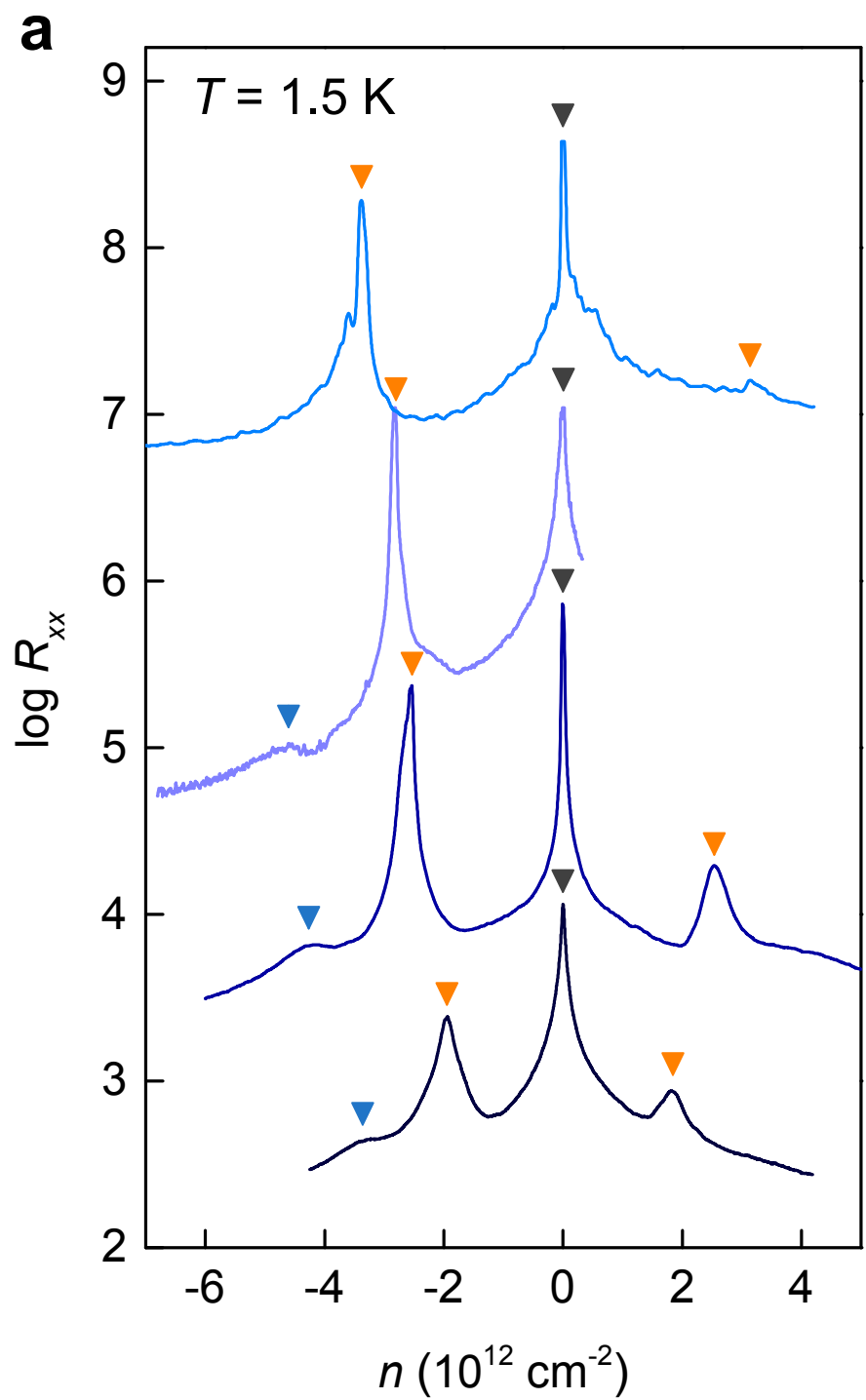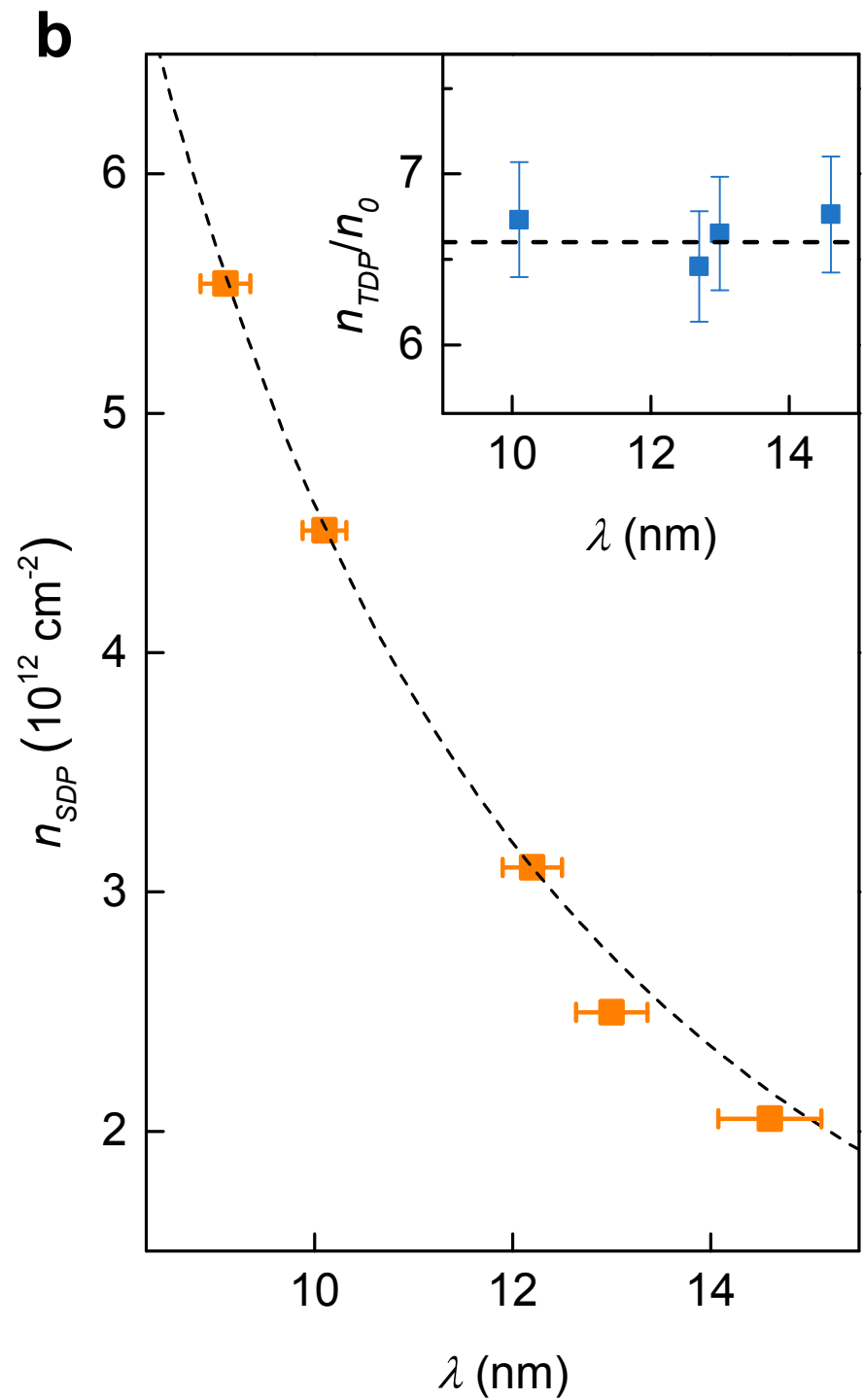

Fig. 2, G. Chen *et al.*

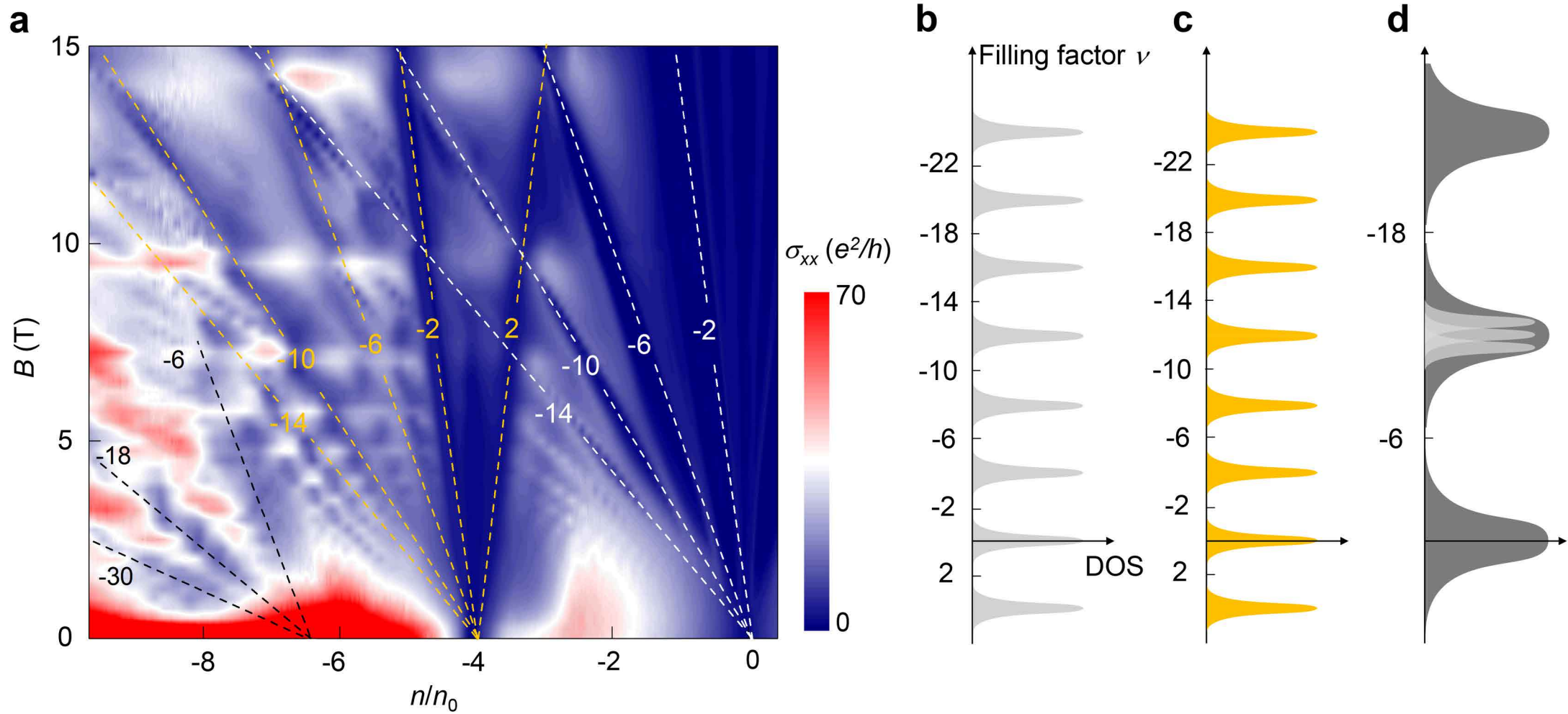

Fig. 3, G. Chen *et al*.

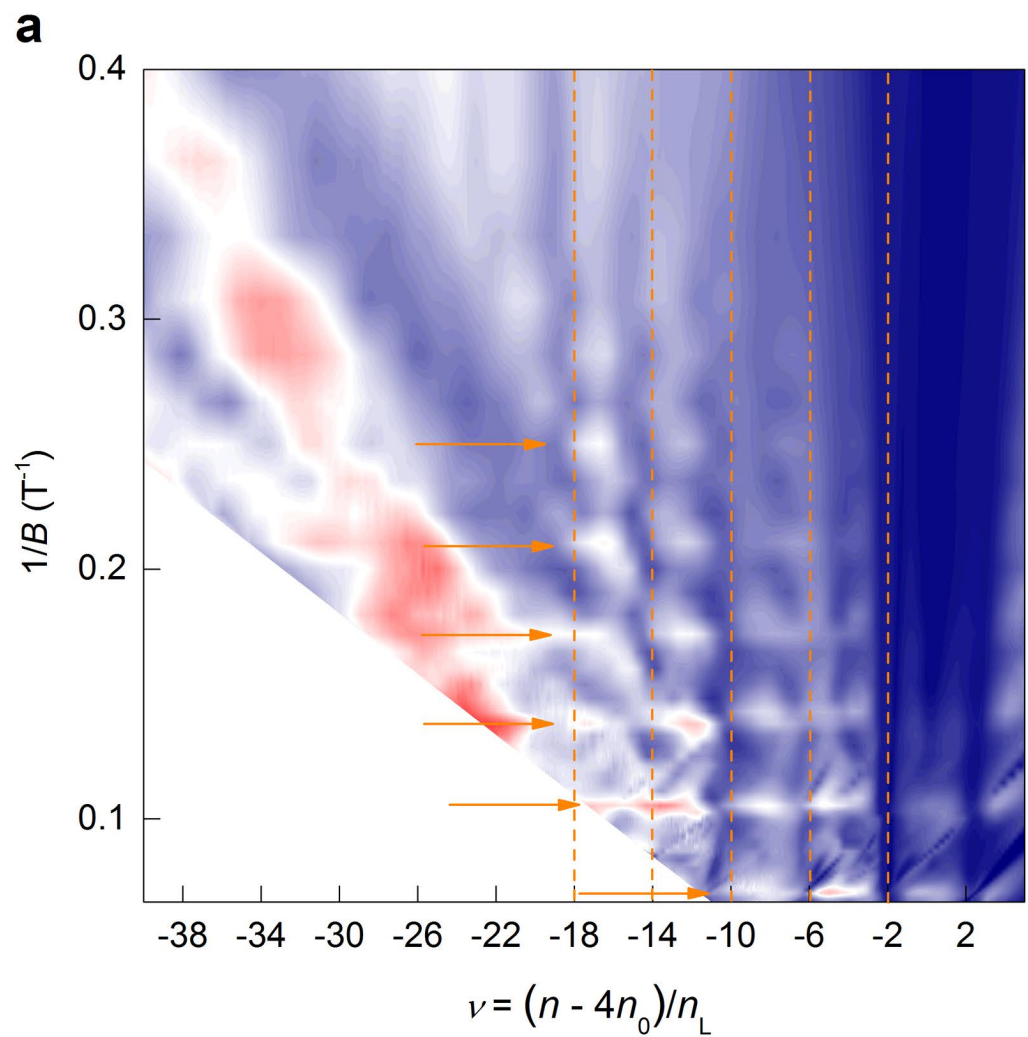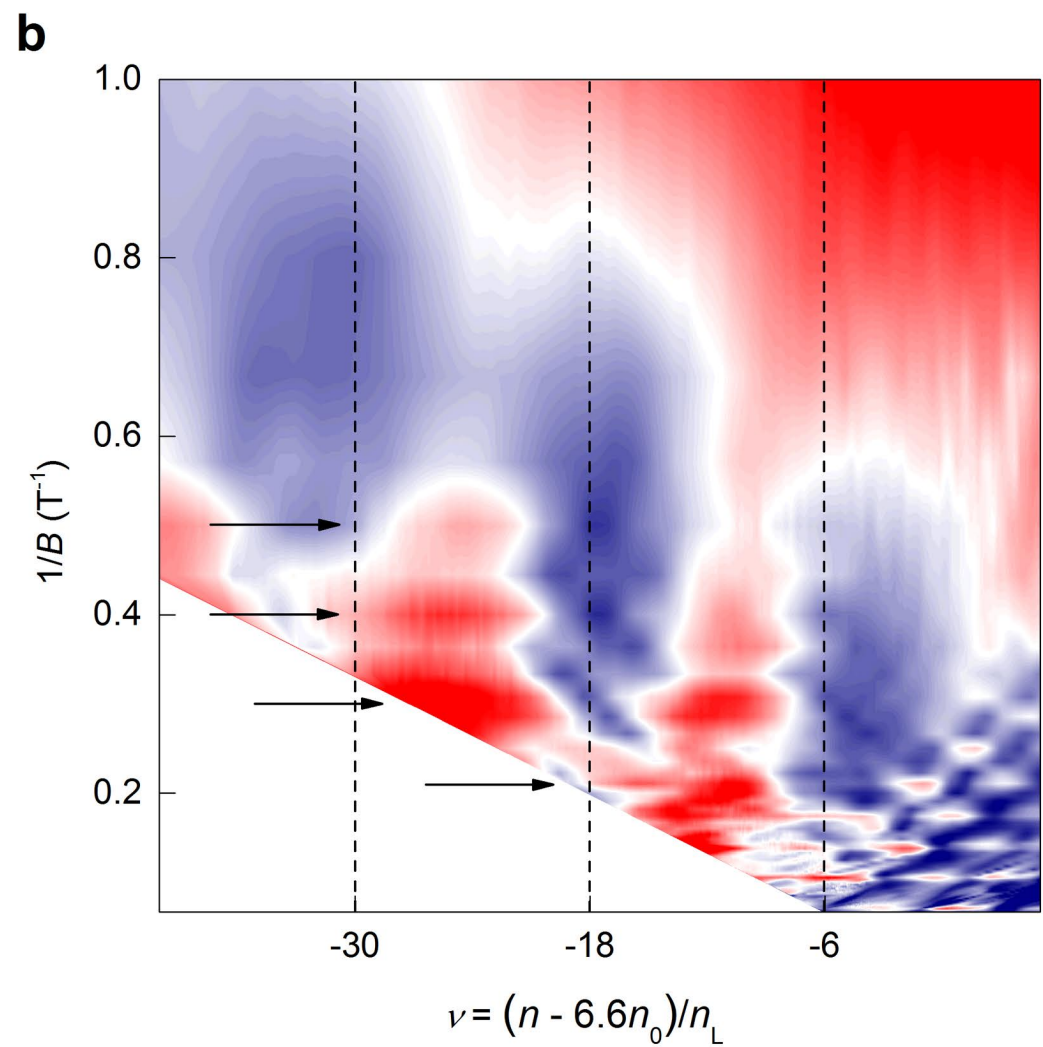

Fig. 4, G. Chen *et al*.

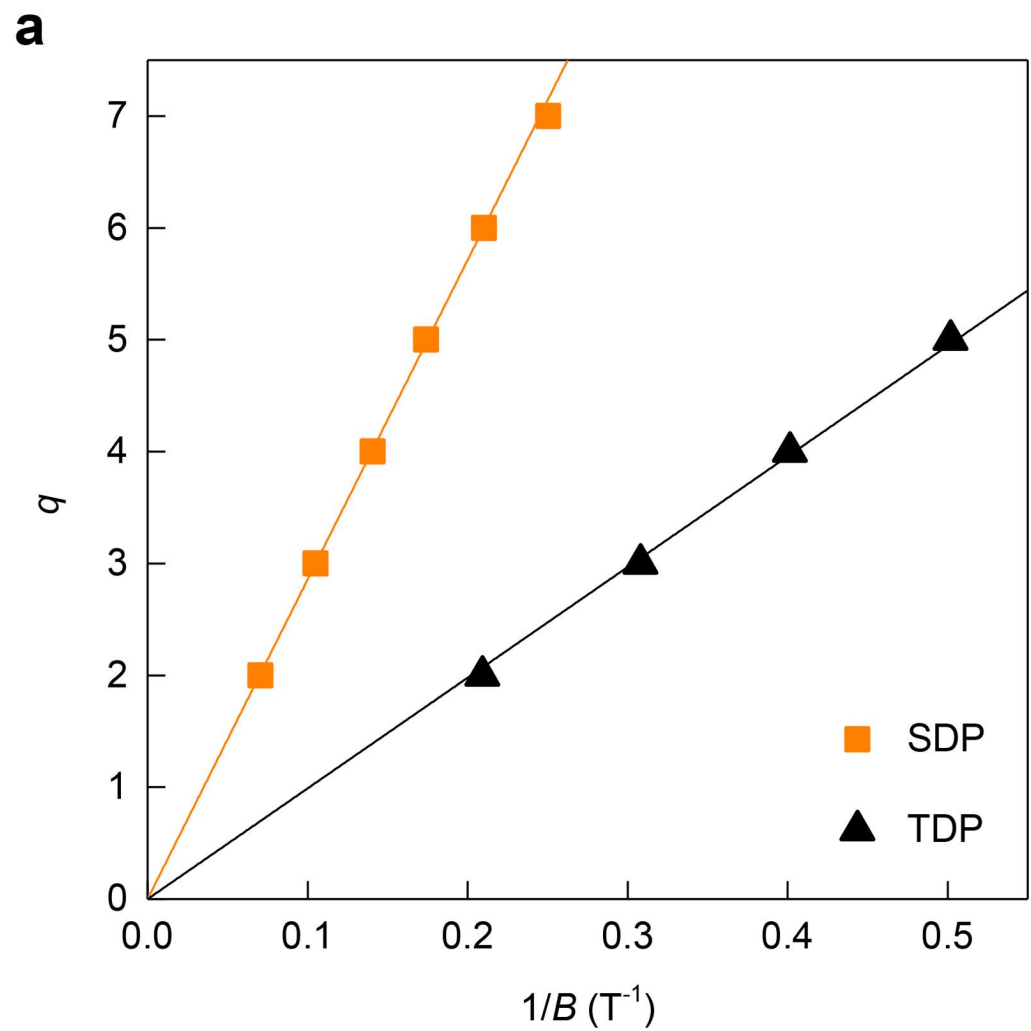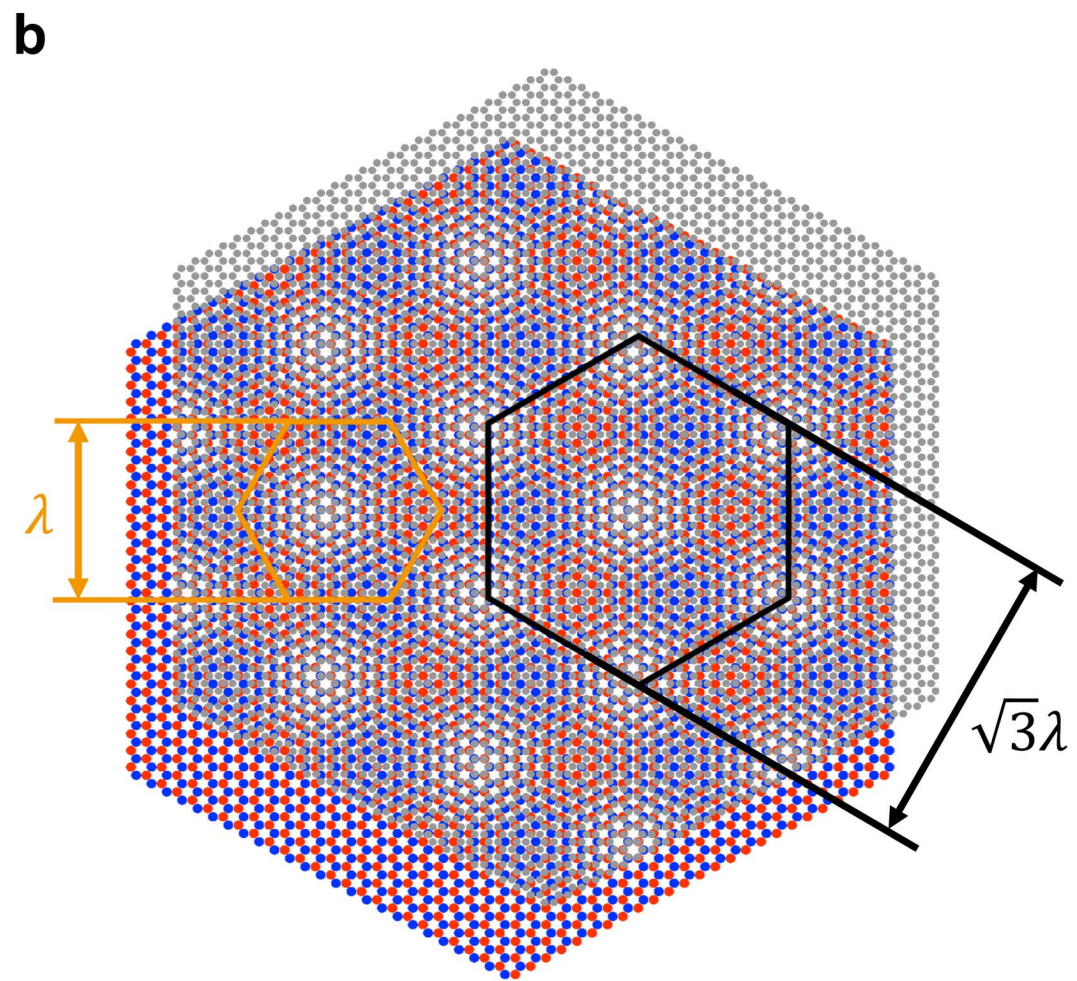

Fig. 5, G. Chen *et al*.